\pdfoutput=1

\documentclass[11pt]{article}

\usepackage[]{acl}

\usepackage{times}
\usepackage{latexsym}

\usepackage[T1]{fontenc}

\usepackage[utf8]{inputenc}

\usepackage{microtype}

\usepackage{inconsolata}

\usepackage{graphicx}

\usepackage{amsmath}
\usepackage{subcaption}
\usepackage{xspace}
\usepackage{url}
\usepackage{multirow}
\usepackage{makecell}
\usepackage{xcolor}         
\usepackage{booktabs}
\usepackage{colortbl}
\usepackage{graphicx}
\usepackage{listings}
\usepackage{float}
\usepackage[para,online,flushleft]{threeparttable}
\newcommand{\ie}{\emph{i.e.,}\xspace}
\newcommand{\eg}{\emph{e.g.,}\xspace}

\newlength\savewidth\newcommand\shline{\noalign{\global\savewidth\arrayrulewidth\global\arrayrulewidth 1pt}\hline\noalign{\global\arrayrulewidth\savewidth}}

\lstdefinestyle{prompt}{
  basicstyle=\footnotesize\ttfamily,
  columns=fullflexible,
  breaklines=true,
  frame=none,
  extendedchars=true,
  escapechar=@,
  literate={á}{{\'a}}1 {ã}{{\~a}}1 {é}{{\'e}}1 {£}{{\pounds}}1 {–}{{-}}1 {’}{{'}}1,
  frame=lines
}

\definecolor{mygray}{gray}{0.9}

\title{Towards Storage-Efficient Visual Document Retrieval: An Empirical Study on Reducing Patch-Level Embeddings}

\author{
 Yubo Ma$^{1}$, Jinsong Li$^{2}$, Yuhang Zang$^{2\ast}$, Xiaobao Wu$^{1}$, Xiaoyi Dong$^{2}$, 
 \\ \textbf{Pan Zhang$^{2}$, Yuhang Cao$^{2}$, Haodong Duan$^{2}$, Jiaqi Wang$^{2}$, Yixin Cao$^3$ , Aixin Sun$^1$} \\
 $^1$ Nanyang Technological University
 $^2$ Shanghai Artificial Intelligence Laboratory \\
 $^3$ Institute of Trustworthy Embodied AI, Fudan University \\
\texttt{yubo001@e.ntu.edu.sg}\\
}

\begin{document}
\maketitle
\renewcommand{\thefootnote}{\fnsymbol{footnote}}
\footnotetext[1]{Corresponding Author}
\begin{abstract}
Despite the strong performance of ColPali/ColQwen2 in Visualized Document Retrieval (VDR), it encodes each page into multiple patch-level embeddings and leads to excessive memory usage. This empirical study investigates methods to reduce patch embeddings per page at minimum performance degradation. We evaluate two token-reduction strategies: \textit{token pruning} and \textit{token merging}. Regarding token pruning, we surprisingly observe that a simple random strategy outperforms other sophisticated pruning methods, though still far from satisfactory. Further analysis reveals that pruning is inherently unsuitable for VDR as it requires removing certain page embeddings without query-specific information. Turning to token merging (more suitable for VDR), we search for the optimal combinations of merging strategy across three dimensions and develop Light-ColPali/ColQwen2. It maintains 98.2\% of retrieval performance with only 11.8\% of original memory usage, and preserves 94.6\% effectiveness at 2.8\% memory footprint. We expect our empirical findings and resulting Light-ColPali/ColQwen2 offer valuable insights and establish a competitive baseline for future research towards efficient VDR.

\end{abstract}

\section{Introduction}
\label{sec:intro}

Visualized Document Retrieval (VDR) matches queries to relevant documents by leveraging their visual representations. Unlike conventional retrieval systems, where raw text must be parsed before indexing, VDR captures documents as images (screenshots) and encodes them into embeddings using Large Vision-Language Models (LVLMs). This approach preserves layout structures and visual elements, enabling retrieval in a what-you-see-is-what-you-get manner. As a result, VDR achieves superior retrieval accuracy and demonstrates strong potential across various applications~\cite{cho2024m3docragmultimodalretrievalneed, chen2025loracontextualizing}.

\begin{figure}
    \centering
    \includegraphics[width=0.8\linewidth]{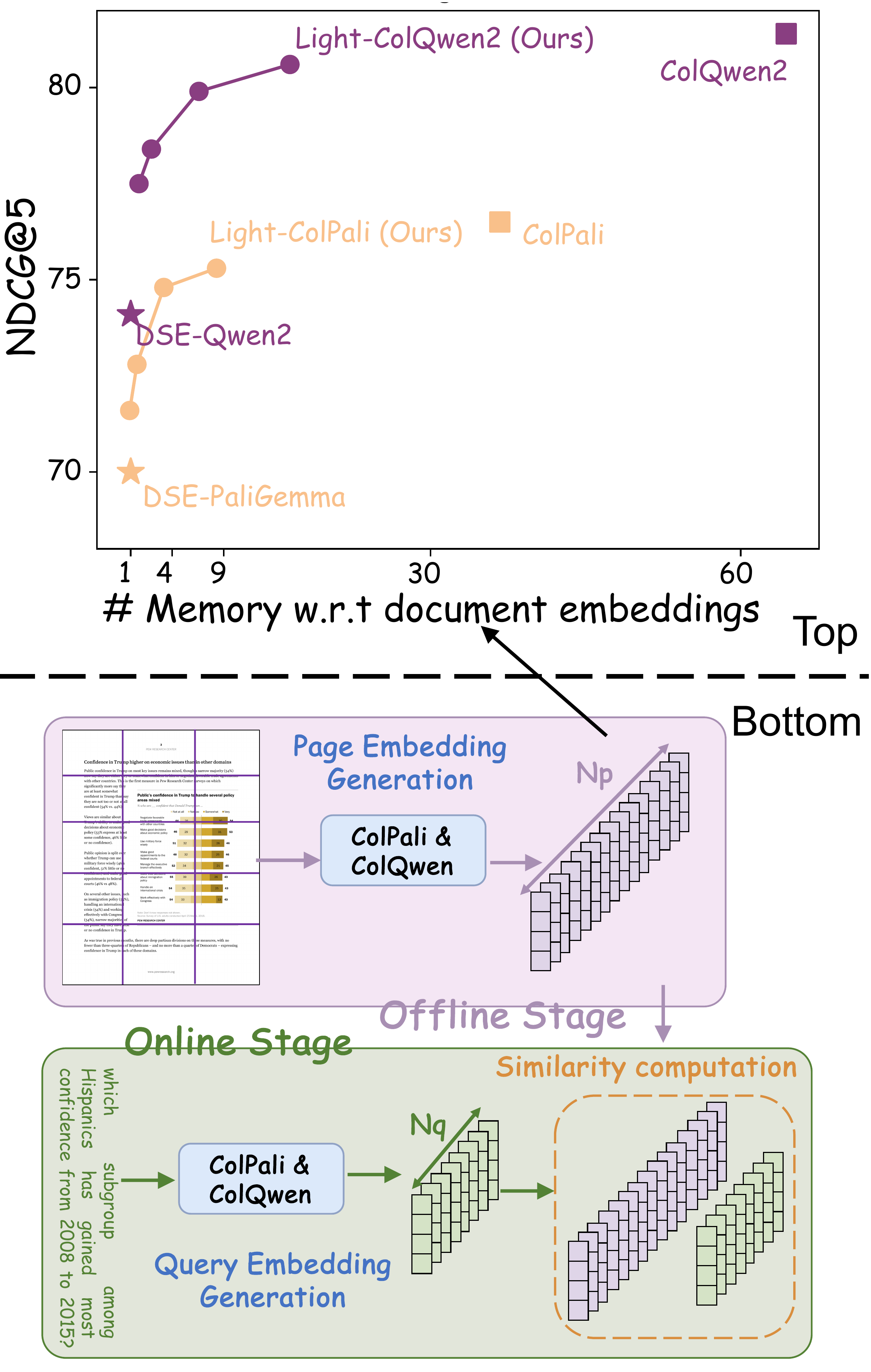}
    \caption{\textbf{Top}: The \textit{relative} memory consumptions for embedding storage of different VDRs.
    Our simple yet effective approach, Light-ColPali/ColQwen2, retains most of the performance but with significantly reduced memory cost. \textbf{Bottom}: The diagram of VDR equipped with ColPali/ColQwen2 retriever. It encodes each page into $N_p$ patch-level embeddings and thus incurs prohibitive memory cost.}
    \label{fig:top_figure}
\end{figure}

The state-of-the-art visualized document retriever, ColPali/ColQwen2~\cite{faysse2025colpali}, represents a significant advancement in this field. As shown in Figure~\ref{fig:top_figure}, ColPali/ColQwen2 encodes each document page as $N_p$ patch-level embeddings during the offline stage and saves them for online computation. While the excessive number of patch embeddings enables the perceiving of fine-grained details (which is particularly important for document-related images), it introduces substantial memory footprints and computational overhead in both offline indexing storage and online similarity computation. For example, a medium-sized document with 50 pages requires about 10 MB memory for embedding storage~\footnote{ColQwen2 divides each page into 768 tokens, each represented by a 128-dimensional vector. Stored as 16-bit floats, it requires 50*768*128*16 bits = 9.6 MB per document.}. This substantial memory footprint presents a bottleneck for scalability and practical deployment of VDR systems under real-world scenarios.

In this work, we present an in-depth analysis of the storage-efficient visualized document retriever, exploring \textit{\textbf{how to reduce each page's patch embedding number with minimal performance degradation}}. We consider two common token-reduction approaches, \ie token pruning~\cite{Chen2024AnII} and token merging~\cite{clavié2024reducingfootprintmultivectorretrieval}, respectively. In terms of token pruning, we investigate multiple pruning strategies in Section~\ref{sec: pruning} and aim to retain only the high-informative patch embeddings. Even though token pruning works to some extent, it can not reduce the embedding numbers by orders of magnitude without significant performance drops. More embarrassingly, we observe that the most simple strategy, \ie random pruning, performs better than other carefully designed strategies. A deeper analysis of this observation reveals that (1) the informativeness of patch embedding is highly conditioned on the queries, which are unknown and unpredictable during the offline indexing stage. (2) the patch embeddings can be grouped and, accordingly, are prone to be dropped by the group under some specific criteria. The above two reasons make it impractical to decide which embeddings should be pruned without access to the queries. Therefore, we claim that pruning-related strategies are inappropriate under VDR settings.

In Section~\ref{sec: merging}, we investigate token merging strategies across three critical dimensions: (1) merging approaches, (2) fine-tuning applicability, and (3) merging locations. Our analysis reveals that similarity-based clustering marginally surpasses spatial-oriented pooling in effectiveness, and resource-efficient fine-tuning (about 72 A100-GPU hours) significantly mitigates the performance gap between retrievers with and without merging. Additionally, we observe that late-stage merging (\ie at the final layer of ColPali/ColQwen2) optimally preserves information and minimizes performance degradation. Building upon these insights, we establish a simple yet effective baseline, named Light-ColPali/ColQwen2, for patch-level embedding reduction in VDR systems. Comprehensive evaluations across three benchmarks~\cite{faysse2025colpali, yu2024visragvisionbasedretrievalaugmentedgeneration, ma2024mmlongbenchdoc} demonstrate that Light-ColPali/ColQwen2 approach maintains comparable performance while achieving orders-of-magnitude reduction in patch storage requirements. Notably, as illustrated in Figure~\ref{fig:top_figure}, it retains 98.2\% of NDCG@5 scores with only 11.8\% of original memory footprint and maintains 94.6\% effectiveness at just 2.8\% memory footprint.
\section{Related Work} 

\textbf{Visualized Document Retriever} shares a similar architecture with text-based dense retrievers~\cite{karpukhin-etal-2020-dense, colbert}, but leverages LVLMs~\cite{wang2024qwen2vlenhancingvisionlanguagemodels, beyer2024paligemmaversatile3bvlm} for OCR-free document understanding. It primarily divides into two approaches: (1) \textit{Page-level embedding retrievers} (DSE;~\citealp{ma-etal-2024-unifying}) encode entire pages and queries into single embeddings; (2) \textit{Patch-level embedding retrievers} (ColPali/ColQwen2;~\citealp{faysse2025colpali}) generate multiple patch-level embeddings per page and token-level embeddings per query. While patch-level retrievers offer finer granularity and superior performance, they demand prohibitive computational resources for both offline indexing and online retrieval. This work addresses this limitation by developing methods to reduce embedding numbers in patch-level retrievers.

\noindent \textbf{Token-reduction about LVLM} has been intensively explored to improve LVLMs' generation (\ie next-token prediction) efficiency. Current approaches fall into three categories: (1) \textit{Pruning strategies}~\cite{Liang2022NotAP, Chen2024AnII, xing2024pyramiddropacceleratinglargevisionlanguage, zhang2025sparsevlmvisualtokensparsification} eliminate low-information tokens based on importance ranking; (2) \textit{Merging strategies}~\cite{kong2022spvit, bolya2023token, TokenPacker} combine similar tokens into compressed embeddings; (3) \textit{Hybrid strategies}~\cite{shang2024llavaprumergeadaptivetokenreduction, yang2024visionziplongerbetternecessary,wu2025sailing} integrate pruning and merging, usually by preserving high-informative tokens while merging redundant ones. In LVLM generation, the focus is on minimizing response latency and FLOPs given specific instructions. In document retrieval, the goal is to reduce the memory footprint of embeddings, without query access but with less concern for latency or FLOPs during indexing. These distinctions relax constraints on token merging, enabling late-stage compression and more computationally-intensive merging strategies. However, the absence of queries precludes query-conditioned pruning or merging approaches.

\noindent \textbf{Lightweight Document Retriever} has been explored to address the challenge of large-scale embeddings with two orthogonal approaches: (1) \textit{Dimension Reduction}. ColBERTv2~\cite{santhanam-etal-2022-colbertv2} employs product quantization~\cite{IVFPQ} to reduce the size of each embedding from 768 to 128 dimensions. This design is inherited by ColPali~\cite{faysse2025colpali} with a simpler projection layer. (2) \textit{Token Reduction}: \citet{clavié2024reducingfootprintmultivectorretrieval} introduces the concept of TokenPooling and explores merging strategies for text-based retrievers. A recent blog by ColPali's author~\cite{tokenpooling-colpali} further extends this to visualized document retrievers. Following their work, our Light-ColPali/ColQwen2 shares very similar merging approaches from the posterior perspectives. However, our work advances this field by conducting a systematic empirical study both on pruning and merging strategies. Beyond simply proposing a merging strategy, our analysis reveals the limitations of pruning (under retrieval settings) and identifies the optimal combination for merging. Moreover, our experiments demonstrate the effectiveness of fine-tuning. Compared to the results reported in~\citet{tokenpooling-colpali}, our fine-tuned Light-ColPali/ColQwen2 presents stronger performance with significantly higher reduction ratios.

\section{The Research Problem} 

\noindent \textbf{ColPali/ColQwen2}. We briefly review the mechanism of ColPali/ColQwen2~\cite{faysse2025colpali} in Figure~\ref{fig:top_figure}. Given query $q$ with $N_q$ tokens and image-formatted document $p$ with $N_p$ patches, ColPali/ColQwen2 encodes them as token-level embeddings $E_q = [e_q^1, ..., e_q^{N_q}] \in R^{N_q \times d}$ and patch-level $E_p  = [e_p^1, ..., e_p^{N_p}] \in R^{N_p \times d}$ into unified embedding space using the LVLM backbone. The relevance score between $q$ and $p$, denoted as $s(q, p)$, is computed by (1) identifying the most similar patch embedding in $p$ for each token in $q$ and (2) summing the similarity scores across all tokens:

\begin{align*}
    & s_j = \texttt{maxsim}(q_j, p) = \max_i {e_p^i}^T e_q^j \\
    & s(q, p) = \sum_j s_j
\end{align*}

In practice, a corpus $C$ of documents is collected and encoded as $E_C \in R^{N_C \times N_p \times d}$ during the offline stage. During the online retrieval stage, when a query $q$ is received and encoded, the top-k most relevant documents are retrieved.

\noindent \textbf{Task Definition}. We notice that each query or page corresponds to multiple token- or patch-level embeddings. In original ColPali/ColQwen2, $N_p$ approximately equals to number of patches determined by the visual encoder in LVLMs, which are 1024 for ColPali and 768 for Qwen2-VL in default. In this work, we investigates various token reduction strategies and produce compressed embeddings $E_p' \in R^{N_p' \times d}$ for each page, where $N_p' \ll N_p$. To this end, we explore two token-reduction strategies, \textit{pruning} and \textit{merging}, in the following sections.

\section{Token Pruning: An Ineffective Strategy}
\label{sec: pruning}

Given patch embeddings $E_p$ for each document page, a natural approach is to retain $N_p'$ embeddings and prune the remaining $(N_p - N_p')$. In this section, we explore three pruning strategies and observe that their performance collapses when reducing embeddings by orders of magnitude. More embarrassingly, the simplest random pruning outperforms other carefully-designed strategies. Further analysis reveals that ColPali's embeddings cluster in groups, while their relevance to different queries is highly unpredictable. These findings highlight the limitations of pruning strategies and underscore the feasibility and necessity of merging strategies under VDR settings.

\subsection{Three Pruning Strategies}
\label{sub_sec: pruning_strategy}

We evaluate three pruning strategies as follows:

\noindent \textbf{Random}: For each $E_p$, we randomly drop $(N_p - N_p')$ embeddings.

\noindent \textbf{Score-oriented}: Recall that ColPali/ColQwen2 measures the query-page relevance by maximum-similarity approach, \ie considering the most similar patch embeddings $e_p^i \in E_p$ with $e_q^j \in E_q$  for each token in $q$. Accordingly, we denote the \textit{response potential} of each patch $p_i \in p$ on query $q$ as its maximum similarities with any token $q_i \in q$, \ie $r_p^i(q) = \max_j {e_p^i}^T e_q^j$. However, the key bottleneck for token-reduction in VDR is exactly that the query $q$, and the associated $r_p^i(q)$, is unknown when we prune $E_p$ at the offline stage. To ensure the performance preservation on any potential $q$, we prompt LVLMs to generate a set of possible queries $Q_p$ given each document page as detailed in Appendix~\ref{appendix:syn_queries}. Then we approximate the response potential on any queries as the maximum values on this sampled set $Q_q$: $r_i^p = \max_{q \in Q_p} r_p^i(q)$. We view patches with low $r_i^p$ values as unimportant for any queries and prune them at priority.

\noindent \textbf{Attention-oriented}: Recall that the common pruning strategies in LVLM's generation~\cite{Chen2024AnII, yang2024visionziplongerbetternecessary} measure the token's importance by their received attentions from other tokens in Transformer layers. We employ this strategy and rank the patch embeddings in $E_p$ by the received attentions (of corresponding tokens in last LVLM layer) from the \texttt{[EOS]} token. We prune embeddings with less attentions at priority.

\begin{figure}
  \centering
  \begin{subfigure}[b]{0.4\textwidth}
    \includegraphics[width=\textwidth]{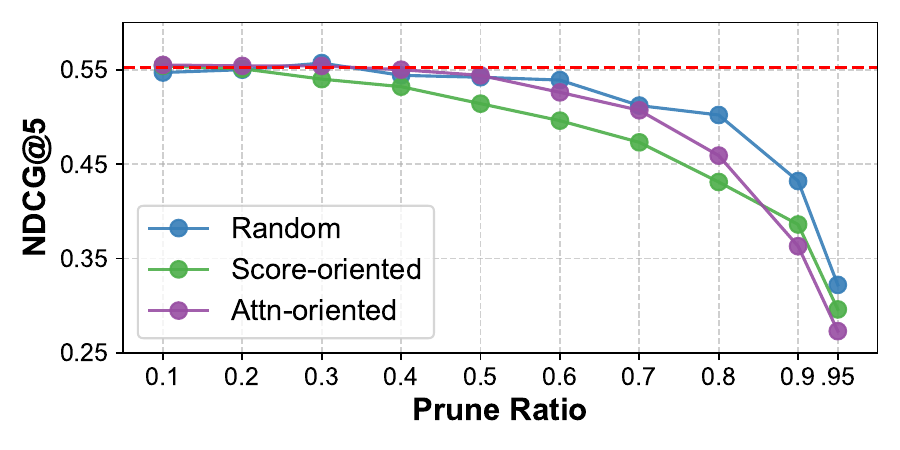}
    \caption{DocVQA}
    \label{fig:image1}
  \end{subfigure}
  \hfill
  \begin{subfigure}[b]{0.4\textwidth}
    \includegraphics[width=\textwidth]{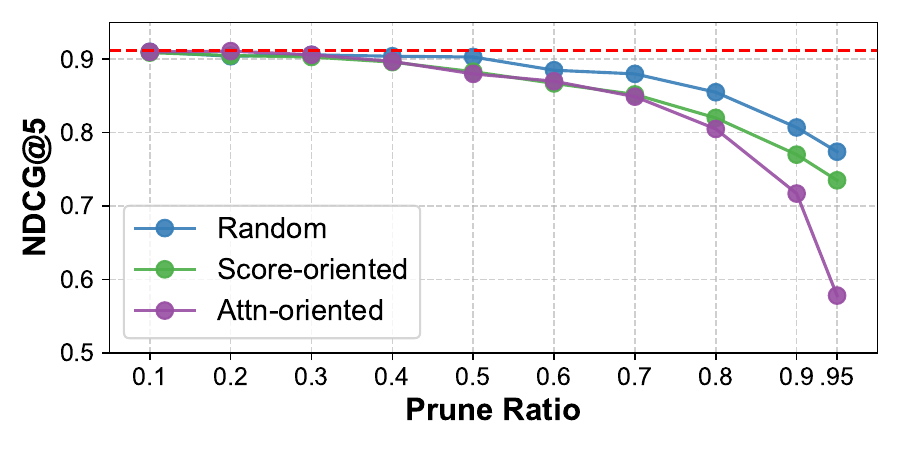}
    \caption{InfoVQA}
    \label{fig:image2}
  \end{subfigure}
  \caption{Retrieval performance v.s. pruning ratio across three different pruning strategies.}
  \label{fig:pruning ratio}
\end{figure}

\subsection{Random Dropping: A Strong SOTA}

We evaluate the pruning strategies above on two representative datasets, DocVQA~\cite{Mathew2020DocVQAAD} and InfoVQA~\cite{Mathew2021InfographicVQA}, from the ViDoRE~\cite{faysse2025colpali} benchmark. The embeddings $E_p$ are generated using the official ColQwen2 checkpoints~\footnote{\url{https://huggingface.co/vidore/colqwen2-v1.0}} and pruned with varying pruning ratios ($1-N_p'/N_p$). As illustrated in Figure~\ref{fig:pruning ratio}, all three strategies maintain their NDCG@5 scores when the pruning ratio is below 0.2, and present slight drop (< 2\% absolute score) for ratios below 0.5. However, more aggressive pruning ratios result in significant performance drop. The best-performing strategy retains only 78.3\% / 88.5\% of its original score at 0.9 pruning ratio and 58.3\% / 84.9\% at 0.95 ratio, which is far from satisfactory. These results demonstrate that none of the three pruning strategies achieve effective token reduction by orders of magnitude. Moreover, we surprisingly observe that the simplest random pruning outperforms the other two strategies, especially when the pruning ratio is above 0.5. At 0.95 pruning ratio, it surpasses the score-oriented strategy by 3.9\% and the attention-oriented strategy by 19.6\% in absolute score on InfoVQA dataset.

\subsection{Analysis}

\begin{figure}
  \centering
  \begin{subfigure}[b]{0.4\textwidth}
    \includegraphics[width=\textwidth]{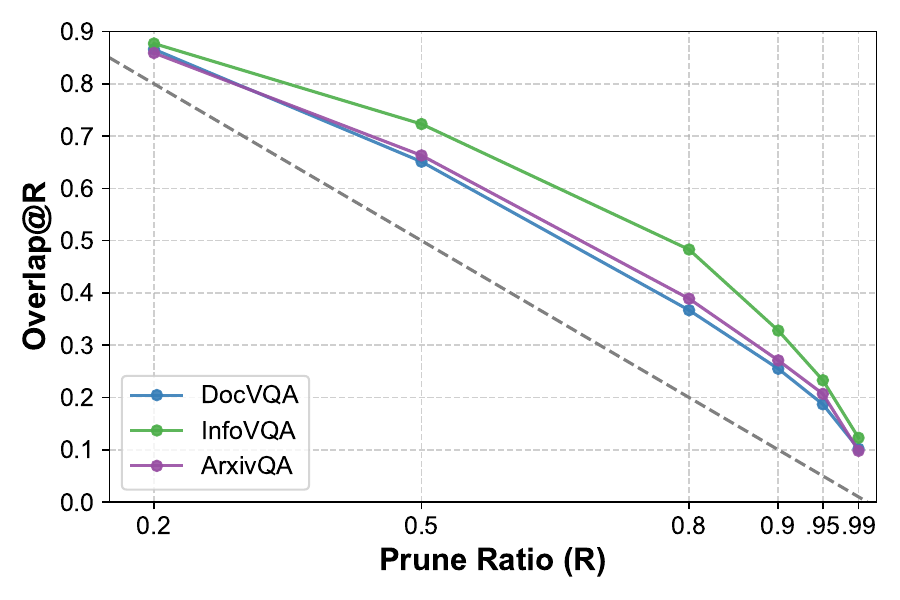}
    \caption{The activated patches overlap of two queries under different pruning ratios.}
  \end{subfigure}
  \hfill
\begin{subfigure}[b]{0.4\textwidth}
    \includegraphics[width=\textwidth]{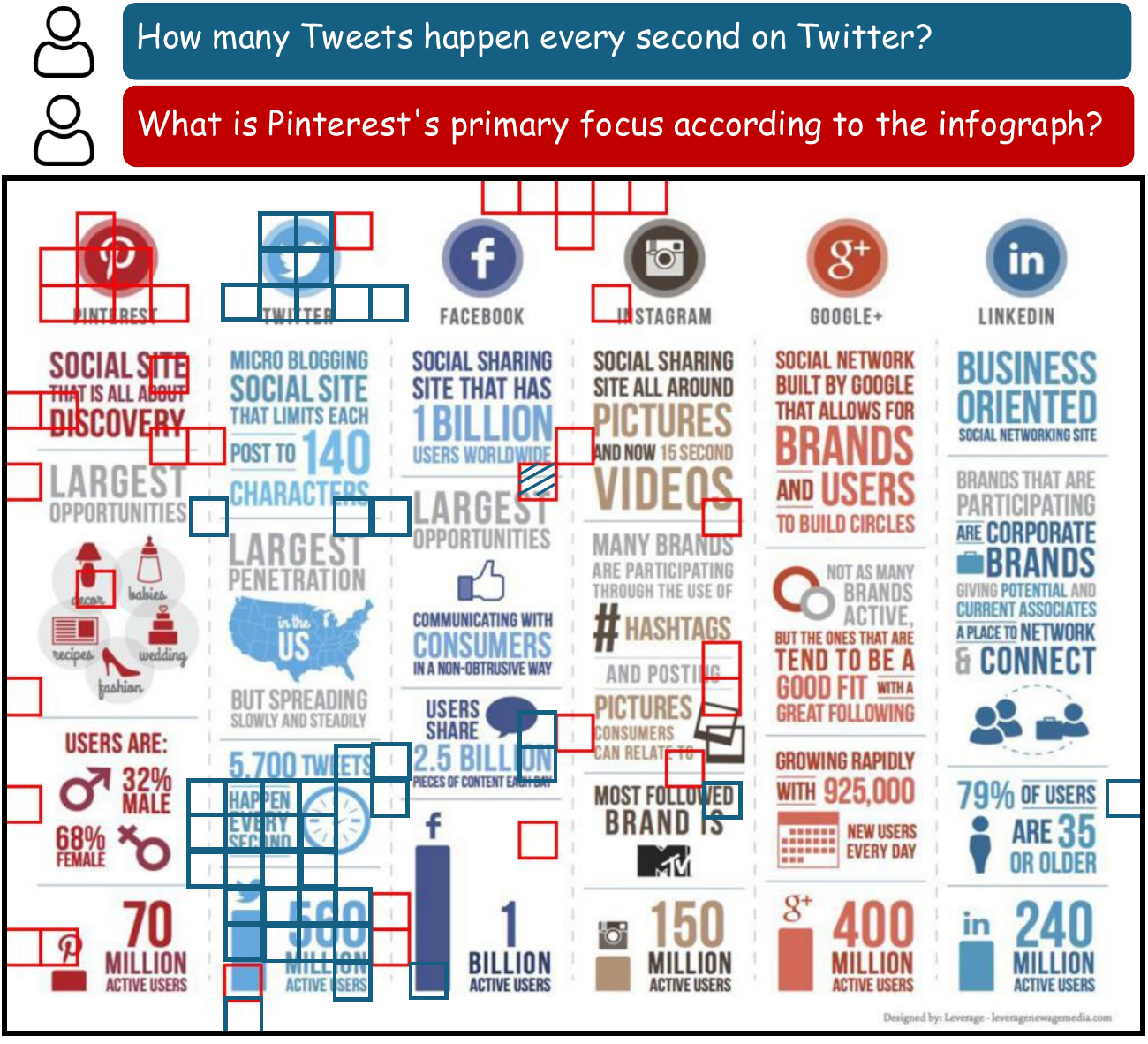}
    \caption{A representative case. The activated patches given different queries are colored in red and blue, respectively. The only shared patch is hatched.}
  \end{subfigure}
  \caption{The triggered patches of the identical page vary with the queries.}
 \label{fig: prune_analysis}
\end{figure}

We investigate the mechanism behind the counter-intuitive observation that random pruning performs best. We attribute it to two key reasons:

\noindent \textbf{The triggered patches of the identical page vary with the queries.} For a document page $p$, an ideal property in the VDR setting is that the distribution $r_p(q) \in R^{N_p}$ remains consistent across different queries $q \in Q$ (\ie small $E_q[\text{KL}(r_p(q)||E_q(r_p))]$ value). In other words, we expect significant overlap in the patches activated (having high $r_p^i$ values) by different queries. Being the foundation of pruning strategy, this consistency allows us to accurately predict and retain informative patches with the help of sampled/simulated queries during the offline stage. To quantitatively evaluate the consistency, we use the synthesized queries $Q_q$ given each page $p$ in Section~\ref{sub_sec: pruning_strategy} to compute $r_p(q)$. Then we define the patches in $p$ activated by $q$ as those with top-$K\%$ highest $r_p^i(q)$ values, and pairwise compute the overlap of activated patches by two different queries. We show the overlap at different prune ratios (\ie 1-$K\%$) in Figure~\ref{fig: prune_analysis}(a). It reveals that the shared activated patches of two queries are only marginally higher than what would occur by random chance (in dashed diagonal). A case shown in Figure~\ref{fig: prune_analysis}(b) further support this result. Given two different queries, the activated patches on the same page are almost exclusive. Only one patch (out of 736; hatched) responds to both queries.

\noindent \textbf{The patch embeddings are redundant.} We define patches as \textit{redundant} if a group of patches on the page respond to the query to a similar extent. We randomly sample 1000 pages from ViDoRE benchmark and compute their normalized values of response potentials as below.

\begin{align*}
    r^{\text{norm}}_p(q) & = \frac{r_p(q)-\min_j r^j_p(q)}{\max_j r^j_p(q) - \min_j r^j_p(q)}
\end{align*}

\noindent The distributions of $r^{\text{norm}}_p(q)$ shown in Appendix~\ref{appendix:redundant} reveal that 14.2 patches have normalized values above 0.95 and 36.9 patches above 0.9 on average. It demonstrates that the image patches are highly redundant and can be grouped.

The above two reasons inherently limit the effectiveness of pruning strategy under VDR setting where the page embeddings should be pruned offline without access to the queries. Since activated patches are unpredictable but their representations are grouped, key patches regarding some query are prone to be dropped \textbf{by group} according to some specific criterion (like attention- or score-oriented). In such case, they even perform worse than random drop because a group of patches are unlikely to be completely dropped \textbf{by random}. Therefore, we claim that pruning strategy is not appropriate for retrieval scenarios and turn to exploring token merging strategies.
\section{Token Merging: The Choices}
\label{sec: merging}

We turn into another token-reduction strategy, \textit{merging}, towards an efficient document visual retriever. Unlike \textit{pruning} which directly drops some tokens, \textit{merging} consolidates the multiple embeddings into one. This approach is particularly suitable for VDR, where the importance of each embedding is highly undetermined (if not conditioned on specific query). We systematically evaluate the \textit{merging} astrategy through three key aspects towards the recipe for the optimal \textit{merging} strategy as detailed below.

\begin{figure}
    \centering
    \includegraphics[width=\linewidth]{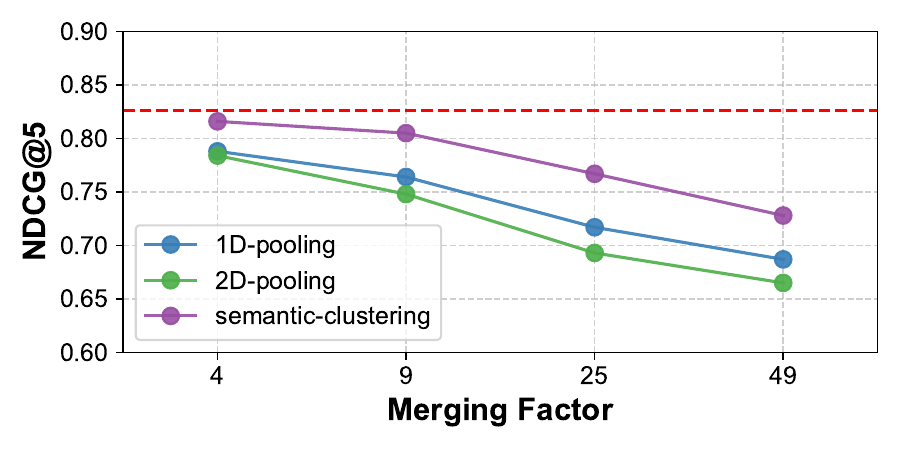}
    \caption{Retrieval performance v.s. merging factor across different merging approaches.}
    \label{fig:merging_method_performance}
\end{figure}

\begin{figure*}
    \centering
    \includegraphics[width=0.95\textwidth]{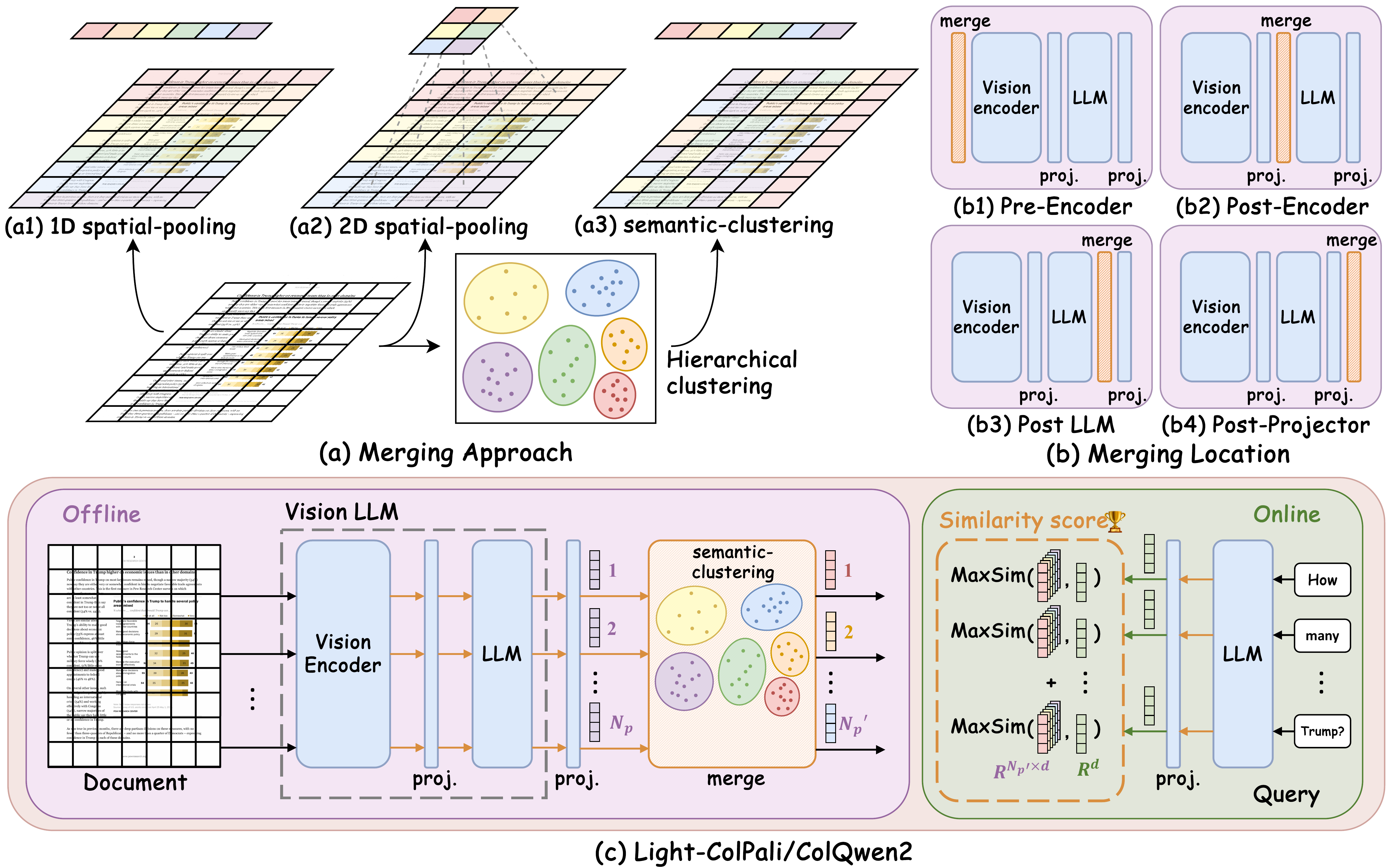}
    \caption{\textbf{(a)}: Three merging approaches. The patches with the same colors are merged into the same embedding. \textbf{(b)}: Three merging locations. Blue blocks represent the original modules in ColPali/ColQwen2. Orange blocks represent the added merging modules. \textbf{(c)}: The architecture diagram  of Light-Colpali/ColQwen2.}
    \label{fig:token_merging}
\end{figure*}

\subsection{Merging Approach}

We follow \citet{clavié2024reducingfootprintmultivectorretrieval} and consider three merging approaches as illustrated in Figure~\ref{fig:token_merging}(a).

\noindent \textbf{1D Spatial-pooling}. In LVLM, images are divided into patches and flattened sequentially. Then their output embeddings are as $R^{N_p \times d}$. To reduce the embeddings from $N_p$ to $N_p'$, the simplest method is to averagely pool every $N_p/N_p'$ embeddings. 

\noindent \textbf{2D Spatial-pooling}. This approach takes into account the spatial structure and semantics of visualized documents. Building on the intuition that adjacent patches often share semantic relationships, 2D-pooling averagely pools embeddings based on their spatial proximity.

\noindent \textbf{Semantic-clustering}. This approach focuses on representation (rather than spatial) proximity. By computing the cosine similarities among the $N_p$  embeddings from ColPali/ColQwen2, we group them into $N_p'$ clusters. Each cluster is then represented by the average of the embeddings within it. Then we conduct hierarchical clustering ~\cite{Murtagh2012AlgorithmsFH} to merge the $N_p$ patch embeddings into $N_p'$ cluster embeddings.

\noindent We evaluate the three merging strategies on six datasets from ViDoRE~\cite{faysse2025colpali} benchmark. We report their average NDCG@5 scores under varying merging factors $N_p/N_p'$ in Figure~\ref{fig:merging_method_performance}. All three merging approaches outperform pruning strategies, with the clustering approach showing particularly strong results. It maintains 97.5\% and 92.6\% relative performance at merging factor 9 and 25, respectively. Such results highlight its effectiveness in maintaining retrieval accuracy even under token reduction by orders of magnitude.

\subsection{Fine-tuning Applicability}

\begin{figure}
    \centering
    \includegraphics[width=0.9\linewidth]{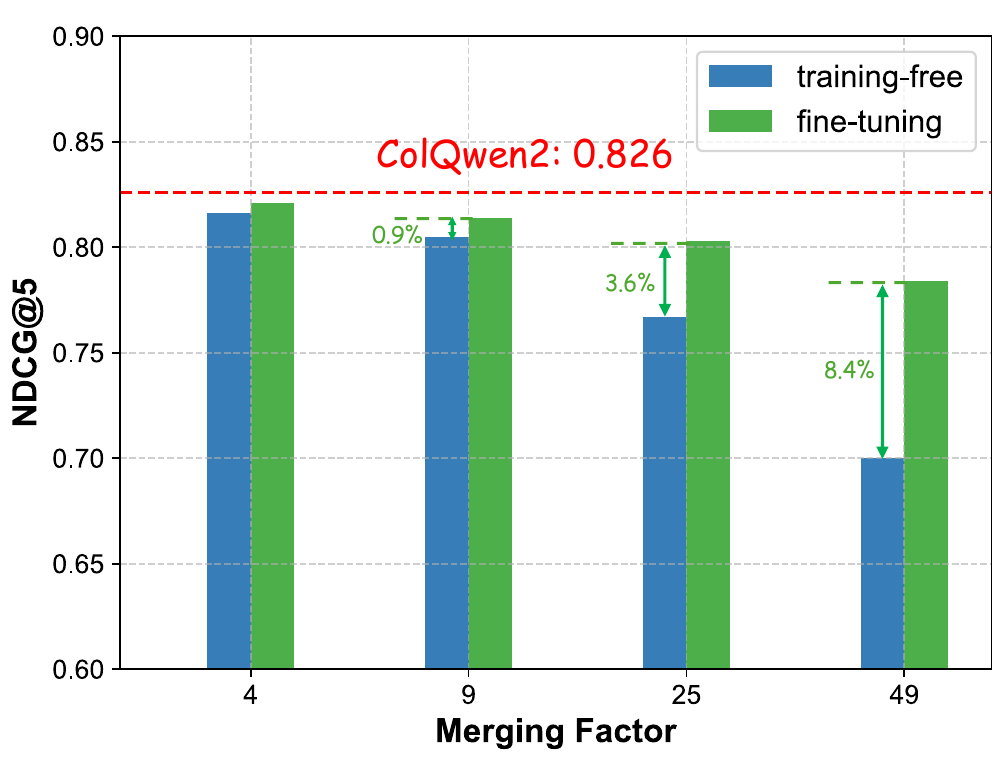}
    \caption{Training-free v.s. fine-tuning retriever with the same merging (clustering) approach. The performance of original ColQwen2 is highlighted in red dash.}
    \label{fig:fine-tuning}
\end{figure}

Above approaches are training-free and serve as plug-and-play modules for the output patch embeddings. While they achieve promising merging ratios without significant performance degradation, we further investigate whether fine-tuning can enhance the performance maintenance. To this end, we compute the relevance score $s(q, p)$ using the merged document embeddings $E_p' \in R^{N_p' \times d}$ during \textbf{BOTH} the training and the inference stage. Results shown in Figure~\ref{fig:fine-tuning} show that fine-tuning retrievers with merged embeddings enhances their perceiving on \textit{blurred} representations and reduces their performance gaps with the original retrievers. This benefit is particularly pronounced at extremely large merging factors. Specifically, at merging factors of 25 and 49 (retaining only 4.6\% and 2.8\% memory cost), fine-tuning recovers 61\% and 67\% of the performance drop (3.6\% and 8.4\% absolutely score gains) caused by training-free. These findings underscore the necessity and effectiveness of fine-tuning in maintaining retrieval performance under aggressive token reduction strategies.

\subsection{Merging Location}

We further explore the optimal location of merging operations within ColPali/ColQwen2. While prior work for efficiency generation~\cite{bolya2023token, Chen2024AnII, yang2024visionziplongerbetternecessary} typically merges tokens in the early layers of LVLMs to reduce FLOPs and response latency, our focus in VDR setting is primarily on the memory footprint of the offline-stored embeddings. This allows us to consider merging operations at later stages, even if FLOPs and latency remain unchanged or increase slightly. Therefore, we explore inserting merging modules at various locations within ColPali/ColQwen2's architecture. As illustrated in Figure~\ref{fig:token_merging}(b), the four options are: (1) Pre-Encoder, (2) Post-Encoder, (3) Post-LLM and (4) Post-projector.

\begin{table}
\small
\centering
\caption{Retrieval performance of different merging locations at merging factor 9.}
\resizebox{\linewidth}{!}{
    \begin{tabular}{l|cccc}
    \toprule
    & \textbf{\makecell{Pre-\\Encoder}} & \textbf{\makecell{Post-\\Encoder}} & \textbf{\makecell{Post-\\LLM}} & \textbf{\makecell{Post-\\Projector}} \\
    \midrule
    \textbf{Info} & 70.2 & 79.5 & 89.7 & 90.4 \\
    \textbf{Doc} & 29.8 & 41.7 & 55.2 & 56.1\\
    \textbf{Arxiv} & 80.0 & 81.9 & 87.6 & 86.7\\
    \textbf{TabF} & 74.1 & 80.8 & 88.6 & 88.8 \\
    \textbf{TAT} & 50.5 & 54.1 & 79.5 & 79.1 \\
    \textbf{Shift} & 49.7 & 54.4 & 85.7 & 87.3 \\
    \midrule
    \textbf{Avg.} & 59.1 & 65.4 & 81.0 & 81.4 \\
    \bottomrule 
    \end{tabular}
    }
    \label{table:merging_location_performance}
\end{table}

We compare the performance of different merging locations at merging factor 9 in Table~\ref{table:merging_location_performance}. We observe that (1) performance significantly improves when the merging operation occurs after LLM module. It demonstrates that token reduction should be performed as late as possible when FLOPs and latency are not the concern, as feeding more visual tokens to the LLM allows for finer-grained perception and more accurate information integration. (2) merging after the final projector yields slightly better performance (0.4\% absolute score) than before it. Since the projector is designed for dimension reduction (\eg from 1536 to 128 for ColQwen2), we hypothesize that clustering algorithms are more effective in low-dimension spaces and thus enable more targeted feature aggregation.
\begin{table*}[t]
    \centering
    \setlength{\tabcolsep}{2.8pt}
    \renewcommand{\arraystretch}{1.4}
    \footnotesize
    \centering
    \caption{The NDCG@5 scores of different visualized document retrievers on three benchmarks. We report their average scores at the most right column, with their relative performance compared with the original ColPali/ColQwen2. We also report their relative memory costs (\textbf{\# Mem}) compared with DSE-Pali/Qwen2.}
    \begin{tabular}{p{2.5cm} cc|c c c c c c|cc|c|c}
        \shline
        & & & \multicolumn{6}{c|}{\textbf{ViDoRE}} & \multicolumn{2}{c|}{\textbf{VisRAG}} &\multirow{2}*{\textbf{MM-LB}} & \multirow{2}*{\textbf{Average}}\\        
        & \makecell{\textbf{Merging} \\ \textbf{Factor}} & \textbf{\# Mem} &  \textbf{Info} & \textbf{Doc} & \textbf{Arxiv} & \textbf{TabF} & \textbf{TAT} & \textbf{Shift} & \textbf{Slide} & \textbf{Chart}& & \\
        \shline
        \rowcolor{mygray}
        \multicolumn{13}{l}{\textit{Base model: Qwen2-VL-2B (original patch number: 768)}} \\ \hline
        DSE-Qwen2 & - & 1.0 & 84.7 & 50.0 & 84.6 & 89.2 & 67.1 & 78.5 & 86.8 & 57.6 & 68.0 & 74.1$_{91.0\%}$ \\
        ColQwen2 & - & 64.4 & 91.5 & 55.4 & 88.0 & 90.5 & 81.1 & 88.5 & 93.4 & 65.8 & 78.6 & 81.4$_{100.0\%}$ \\
        ColQwen2$_{+\text{Pruning}}$ & 9 & 7.6 & 85.6 & 48.3 & 84.0 & 88.3 & 68.6 & 72.5 & 89.3 & 60.3 & 69.0 & 74.0$_{90.9\%}$ \\
        & 49 & 1.8 & 74.7 & 36.3 & 77.1 & 80.5 & 46.7 & 55.9 & 77.3 & 52.8 & 62.3 & 62.6$_{76.9\%}$ \\
        \hline
        Light-ColQwen2 & 4 & 16.4 & 89.5 & 56.6 & 88.6 & 90.2 & 80.5 & 87.1 & 92.9 & 62.9 & 77.0 & 80.6$_{99.0\%}$  \\
        & 9 & 7.6 & 90.4 & 56.1 & 86.7 & 88.8 & 79.1 & 87.3 & 92.2 & 62.0 & 76.2 & 79.9$_{98.2\%}$ \\
        & 25 & 3.0 & 88.9 & 54.6 & 86.4 & 89.3 & 78.7 & 84.4 & 91.0 & 60.4 & 71.9 & 78.4$_{96.3\%}$ \\
        & 49 &  1.8 & 86.9 & 52.6 & 86.5 & 86.8 & 73.5 & 84.5 & 89.7 & 59.6 & 72.8 & 77.0$_{94.6\%}$ \\    
        \shline
        \rowcolor{mygray}
        \multicolumn{13}{l}{\textit{Base model: PaliGemma-3B  (original patch number: 1024)}} \\ \hline
        DSE-Pali & - & 1.0 & 80.1 & 46.0 & 82.0 & 84.1& 61.1 & 70.2 & 84.8 & 54.7 & 67.0 & 70.0$_{91.5\%}$ \\
        ColPali & - & 36.7 & 84.4 & 54.8 & 85.1 & 85.3 & 72.3 & 75.5 & 92.2 & 62.0 & 77.1 & 76.5$_{100.0\%}$ \\
        ColPali$_{+\text{Pruning}}$ & 9 & 4.2 & 81.5 & 50.5 & 82.0 & 84.4 & 61.1 & 67.0 & 90.2 & 59.0 & 69.1 & 71.6$_{93.6\%}$ \\
        & 49 & 0.9 & 72.5 & 35.8 & 70.3 & 72.6 & 40.3 & 44.1 & 79.1 & 50.3 & 61.9 & 58.6$_{76.6\%}$ \\
        \hline
        Light-ColPali & 4 & 9.3 &  82.8 & 53.4 & 84.1 & 86.5 & 72.8 & 72.5 & 91.7 & 60.6 & 73.3 & 75.3$_{98.4\%}$ \\
         & 9 & 4.2 & 82.1 & 54.8 & 83.5 & 84.5 &70.9  & 72.8 & 91.2 & 61.0 & 72.6 & 74.8$_{97.8\%}$ \\
        & 25 & 1.6 & 81.2 & 50.5 & 82.6 & 82.7 & 67.2 & 70.7 & 90.8 & 57.3 & 71.9 & 72.8$_{95.2\%}$ \\
        & 49 & 0.9 & 79.9 & 49.6 & 82.7 & 81.9 & 67.4 & 69.0 & 88.9 & 57.5 & 68.8 & 71.6$_{93.6\%}$ \\
        \shline
	\end{tabular}
	\label{tab:main_results}
\end{table*}

\section{Light-ColQwen2: Effective Storage Reduction on Patch-level Embeddings}

We conduct extensive experiments to identify the optimal merging strategy in Section~\ref{sec: merging}. The key findings are as follows: (1) \textit{Merging Approach}: Merging upon representation similarity (semantic clustering) outperforms spatial proximity (1D- / 2D-spatial pooling). (2) \textit{Merging Location}: Merging at the last stage of retrievers fully leverages the powerful perception capabilities of LVLMs and thus achieves minimal performance drop. (3) \textit{Fine-tuning Applicability}: Incorporating the merging module during training stage significantly reduces the gap compared to the original retrievers, particularly at high reduction ratios. 

Based on these insights, we propose a simple yet effective token-reduction approach for ColPali/ColQwen2, named Light-ColPali/ColQwen2. As illustrated in Figure~\ref{fig:token_merging}(c), it is a token merging strategy which integrates semantic clustering at the latest stage of the pipeline, combined with fine-tuning, to achieve efficient and accurate visualized document retrieval. The simplicity and effectiveness of Light-ColPali/ColQwen2 make it a practical solution for balancing performance and efficiency in visual document retrieval tasks.

\noindent \textbf{Baseline}
We evaluate Light-ColPali/ColQwen2 against three primary baselines. (1) The original ColPali/ColQwen2~\cite{faysse2025colpali} which encodes each patch in the page as one embedding. (2) DSE-Pali/-Qwen2~\cite{ma-etal-2024-unifying} which encodes each page into one embedding. (3) The most effective pruning strategy, random pruning, as introduced in Section~\ref{sub_sec: pruning_strategy}. Toward a fair comparison, all above baselines and our ColPali/ColQwen2 are fine-tuned under the same settings detailed in Appendix~\ref{appendix: training_details}, and are compared in terms of both retrieval performance and memory cost.

\noindent \textbf{Experiment Setup}
We conduct experiments on nine datasets from three benchmarks: ViDoRE~\cite{faysse2025colpali}, VisRAG~\cite{yu2024visragvisionbasedretrievalaugmentedgeneration} and MMLongBench-Doc~\cite{ma2024mmlongbenchdoc} as detailed in Appendix~\ref{appendix_benchmarks}. We follow previous work to use NDCG@5 as the evaluation metric on performance and relative memory cost (compared with DSE) as the metric on efficiency.

\noindent \textbf{Result}
Based on Qwen2-VL-2B~\cite{wang2024qwen2vlenhancingvisionlanguagemodels} and PaliGemma-3B~\cite{beyer2024paligemmaversatile3bvlm}, we show results of different visualized document retrievers on Figure~\ref{fig:top_figure} and Table~\ref{tab:main_results}. The results about Qwen2-VL-7B are supplemented in Appendix~\ref{appendix: more_results}. We observe that (1) ColPali/ColQwen2 achieves superior performance but at the cost of a significantly larger memory footprint compared to DSE. Specifically, ColPali/Qwen2 outperforms DSE by 7.3\% absolute scores on Qwen2-VL-2B and 6.5\% absolute scores on PaliGemma-3B. However, this performance gain comes with a substantial memory overhead requiring 64.4 times and 36.7 times more memory, respectively. This high memory cost imposes significant burdens on both offline indexing and online retrieval and highlights the necessity for a performance-cost balance. (2) Light-ColPali/ColQwen2 achieves a significant reduction in memory footprint while largely preserving performance. For Light-ColQwen2, it maintains 99.0\% of NDCG@5 scores (80.6 out of 81.4) at a merging factor of 4 (\ie retaining only 25.5\% of the memory cost) and 98.2\% of NDCG@5 scores at a merging factor of 9. Even at an extremely large merging ratio, where its memory cost is comparable to DSE (1.8x), Light-ColQwen2 retains 94.5\% relative performance and outperforms DSE by 2.9\% in absolute score gains. Similarly, Light-ColPali maintains 98.4\% and 97.8\% of NDCG@5 scores at merging factors of 4 and 9, respectively. Furthermore, at an extreme reduction ratio of 49 (even lower memory cost than DSE), Light-ColPali retains 93.6\% relative performance and surpasses DSE by 1.6\% in absolute score gains. These results demonstrate that Light-ColPali/ColQwen2 effectively balances memory efficiency and retrieval performance, offering a practical solution for less memory cost without sacrificing significant accuracy. (3) Light-ColPali/ColQwen2 exhibits varying levels of performance preservation across different datasets. For InfoVQA, ArxivQA, TabFQuAD and SlideVQA where documents typically have lower information densities (\eg posters, diagrams), the performance retention is notably higher. In contrast, for datasets like DocVQA, TAT-DQA, and ChartQA where documents are more text-rich and incorporates more information, the performance drop is slightly more obvious. We speculate that the optimal merging factor for each document page highly correlates with its information density. However, how to adaptively adjust the merging factor, both during training and inference stage, remains an open challenge. We leave this as future work.

\begin{table}
    \centering
    \small
    \caption{Time cost of ColQwen2 v.s. Light-ColQwen2 during offline stage. \textbf{Training}: 5 epochs (2310 steps with batch size 128) on 8 A100 GPUs. \textbf{Embed Gen}: 500 page embeddings on single A100 GPU.}
    \begin{threeparttable}
    \begin{tabular}{l|cccc}
        \toprule
         \multirow{2}{*}{\textbf{Model}} & \multicolumn{2}{c}{\textbf{ColQwen2}} & \multicolumn{2}{c}{\textbf{Light-ColQwen2}} \\
         & 2B & 7B & 2B & 7B \\
         \midrule
         \textbf{Training} & 5.6 h & 7.5 h & 9.0 h & 10.5 h \\
         \textbf{Embed Gen} & 1.7 min & 2.1 min & 2.6 min & 3.0 min \\
         \bottomrule
    \end{tabular}
    \end{threeparttable}
    \label{tab: speed}
\end{table}

\noindent \textbf{Time Cost (Offline Stage)} The clustering operation in Light-ColPali/ColQwen2 incurs a modest additional time cost during both model training and embedding generation in the offline stage. As shown in Table~\ref{tab: speed}, it adds 3-3.5 hours to the training time and 0.9 minute to the document embedding generation time. We consider this slight increase in offline latency acceptable given the substantial reduction in memory footprint and the resulting acceleration during the online retrieval stage.
\section{Conclusion}
This work conducts an empirical study into developing efficient visualized document retrievers with minimal memory footprints and performance loss. Through comprehensive experiments, we demonstrate the superior suitability of merging for VDR tasks. Our proposed Light-ColPali/ColQwen2, a simple yet effective merging strategy, achieves significant memory reduction while maintaining promising performance. These findings and the established baseline provide valuable insights for advancing efficient VDR research.

\section*{Limitations}
The primary limitation of this work is the focused scope. We exclusively concentrate on token reduction for minimizing document embedding storage. Alternative aspects for efficient VDR such as dimension reduction, vector quantization, data cleaning and model distillation, remain unexplored in our work. 
However, we emphasize that these techniques are orthogonal to our work and could potentially complement our findings. Future research could integrate these methods with our token reduction explorations to achieve greater efficiency with less performance degradation.

\section*{Acknowledgement}
This work is supported under the RIE2020 Industry Alignment Fund – Industry Collaboration Projects (IAF-ICP) Funding Initiative, as well as cash and in-kind contribution from the industry partner(s). This work is also supported by National Key R\&D Program of China 2022ZD0161600, Shanghai Artificial lntelligence Laboratory.

\bibliography{custom}

\clearpage
\appendix
\section{Details for Token Pruning Experiments}

\subsection{Synthesized Queries}
\label{appendix:syn_queries}

Given a document page, we synthesize multiple queries to explore the possibility to estimate its patch-level potential responses in Section~\ref{sub_sec: pruning_strategy}. Specifically, we prompt Qwen2-VL-7B~\cite{wang2024qwen2vlenhancingvisionlanguagemodels} to generate five more queries as below.

\begin{lstlisting}[caption={The used prompt.}, style=prompt]
Image: <Image>

Given the screenshot of a document/poster, you are asked to generate five question that can be answered by looking at the image.  The questions should be relevant to the content of the image. The questions should be unique and be of varying question types. The output should be formatted as:
1. [Question 1]
2. [Question 2]
3. [Question 3]
4. [Question 4]
5. [Question 5]

Answer: <Answer>
\end{lstlisting}

\subsection{Distribution of Normalized Scores for Response Potential}
\label{appendix:redundant}

\begin{figure}[H]
    \centering
    \includegraphics[width=0.9\linewidth]{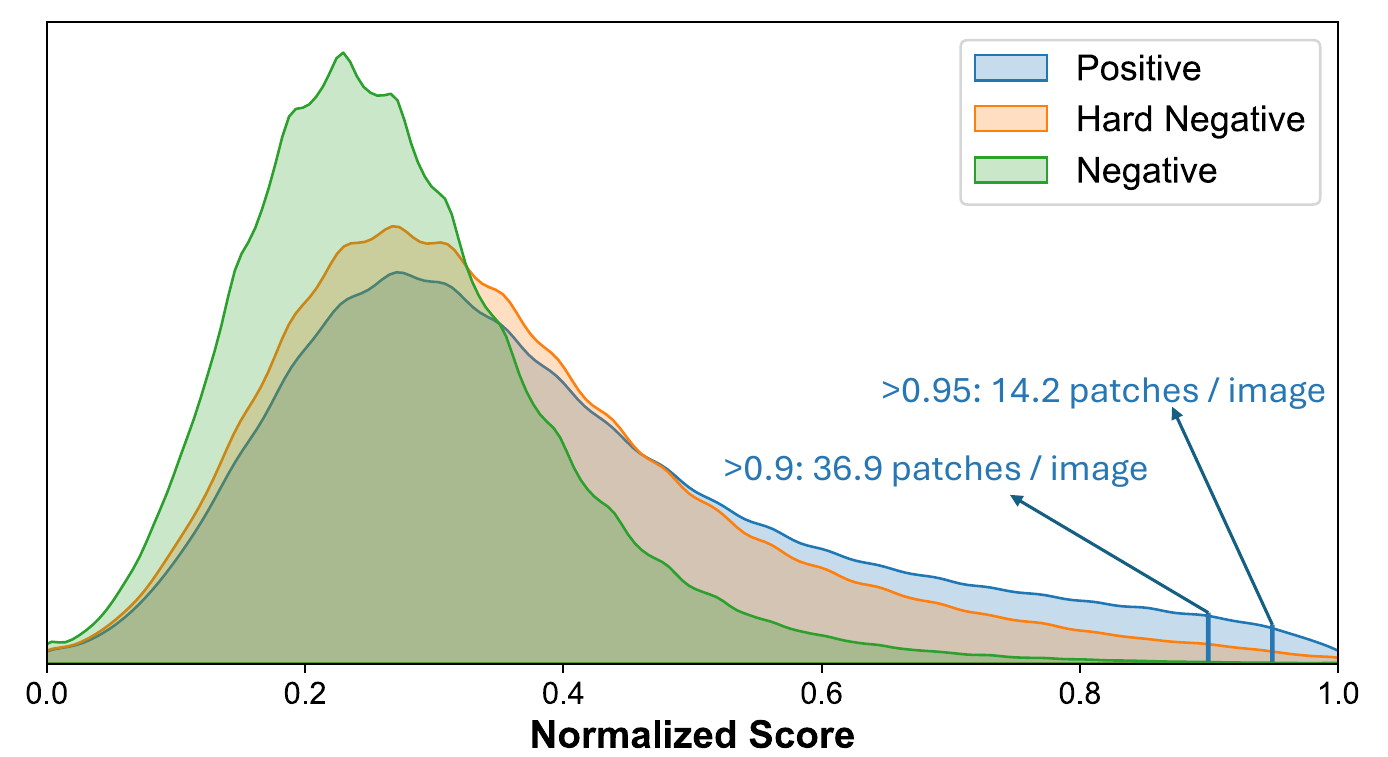}
    \label{fig:kde_plot}
    \vspace{-1em}
\end{figure}
\section{Details for Token Merging Experiments}

\subsection{Benchmarks}
\label{appendix_benchmarks}
We evaluate Light-ColPali/ColQwen2 on nine datasets from three benchmarks as detailed below. All of these three benchmarks are licensed under the Creative Commons license (CC-BY) or other open-source licenses \cite{wu2024akew,wu2024antileak}.

\noindent \textbf{ViDoRE}~\cite{faysse2025colpali}. We select six datasets from ViDoRE: InfoVQA~\cite{Mathew2021InfographicVQA}, DocVQA~\cite{Mathew2020DocVQAAD}, ArxivQA~\cite{li-etal-2024-multimodal-arxiv}, TAT-DQA~\cite{zhu2022towards}, TabFQuAD and Shift Project. Except Shift Project, we remove another four synthesized datasets in ViDoRE because current visualized document retriever has achieved saturated performance on these oversimple datasets.

\noindent \textbf{VisRAG}~\cite{yu2024visragvisionbasedretrievalaugmentedgeneration}. To enhance the evaluation coverage, we additionally select two datasets, ChartQA~\cite{masry2022chartqabenchmarkquestionanswering} and SlideVQA~\cite{SlideVQA2023} from VisRAG. The other datasets in it are not included since they have large overlaps with datasets in ViDoRE.

\noindent \textbf{MMLongBench-Doc}~\cite{ma2024mmlongbenchdoc}. We also incorporate this long-context document understanding dataset in our evaluation. We select the 485 single-page questions as the queries and the screenshots of all PDF pages as document pages. Given a query, note that the retrieved candidate pages are constrained on the ground-truth PDFs. 

\subsection{Training Details}
\label{appendix: training_details}
We fine-tune PaliGemma~\cite{beyer2024paligemmaversatile3bvlm} and Qwen2-VL~\cite{wang2024qwen2vlenhancingvisionlanguagemodels} to reproduce the ColPali/ColQwen2 and DSE-Pali/DSE-Qwen2, respectively. All experiments (including the Light-ColPali/ColQwen2) are based on the ColPali's codebase~\footnote{\url{https://github.com/illuin-tech/colpali}}. For a fair comparison, we train the models on the same training dataset used by the original ColPali which incorporates over 130k queries. The training process lasts for 5 epochs. The batch size is 256 and the learning rate is 5e-4. And we use LoRA~\cite{hu2022lora} with $\alpha$ = 32 and $r$ = 32 on the transformer layers within the language models to reduce the training parameters. We set the temperature coefficient of the InfoNCE loss in DSE as 0.07 and observe a significant performance improvement.

\subsection{More Results on Qwen2-VL-7B}
\label{appendix: more_results}
The results about different document visualized retrievers, with base model Qwen2-VL-7B, are shown in Table~\ref{tab:Supp_results}. 

\begin{table*}[t]
    \centering
    \setlength{\tabcolsep}{2.8pt}
    \renewcommand{\arraystretch}{1.4}
    \footnotesize
    \centering
    \caption{The NDCG@5 scores of different visualized document retrievers on base model: Qwen2-VL-7B.}
    \begin{tabular}{p{2.5cm} cc|c c c c c c|cc|c|c}
        \shline
        & & & \multicolumn{6}{c|}{\textbf{ViDoRE}} & \multicolumn{2}{c|}{\textbf{VisRAG}} &\multirow{2}*{\textbf{MM-LB}} & \multirow{2}*{\textbf{Average}}\\        
        & \makecell{\textbf{Pooling} \\ \textbf{Factor}} & \textbf{\# Mem} &  \textbf{Info} & \textbf{Doc} & \textbf{Arxiv} & \textbf{TabF} & \textbf{TAT} & \textbf{Shift} & \textbf{Slide} & \textbf{Chart}& & \\
        \shline
        \rowcolor{mygray}
       \multicolumn{13}{l}{\textit{Base model: Qwen2-VL-7B}} \\
       \hline
        DSE-Qwen2 & - & 1.0 & 87.3 & 52.3 & 87.9 & 92.3 & 73.0 & 84.8 & 89.6 & 61.8 & 69.6 & 77.6$_{95.0\%}$ \\
        ColQwen2 & - & 36.7 & 91.9 & 56.2 & 89.8 & 90.3 & 86.9 & 82.2 & 93.5 & 65.7 & 79.2 & 81.7$_{100.0\%}$ \\
        \hline
        Light-ColQwen2 & 4 & 9.3 & 91.1 & 55.5 & 90.0 & 91.8 & 81.1 & 85.7 & 93.4 & 64.2 & 78.1 & 81.2$_{99.4\%}$ \\
        & 9 & 4.2 & 91.5 & 56.8 & 88.5 & 92.4 & 80.9 & 87.3 & 93.2 & 63.3 & 76.8 &  81.2$_{99.4\%}$ \\
        & 25 & 1.6 & 90.5 & 54.6 & 89.0 & 91.8 & 79.8 & 84.6 & 91.6 & 61.1 & 77.0 & 80.0$_{98.0\%}$  \\
        & 49 & 0.9 & 89.6 & 52.6 & 88.2 & 89.5 & 76.5 & 81.2 & 90.8 & 58.8 & 72.7 & 77.8$_{95.2\%}$ \\
        \shline
	\end{tabular}
	\label{tab:Supp_results}
\end{table*}

\end{document}